\documentclass[pra,aps,twocolumn,floatfix,superscriptaddress,showpacs]{revtex4-1}
\usepackage{graphicx}
\usepackage{array}
\usepackage{color}
\usepackage{amsmath}
\usepackage{amsxtra}
\usepackage{amstext}
\usepackage{amssymb}
\usepackage{latexsym}
\usepackage{dsfont}

\begin{document}

\title{One qubit and one photon -- the simplest polaritonic heat engine}

\author{Qiao Song}
\affiliation{Quantum Institute for Light and Atoms, School of Physics and Material Science, East China Normal University, Shanghai 200241, P.R. China}
\author{Swati Singh}
\affiliation{Department of Physics and College of Optical Sciences, University of Arizona, Tucson, Arizona 85721, USA}
\affiliation{Department of Physics, Williams College, Williamstown, MA 01267, USA}
\author{Keye Zhang}
\email[Email: ]{kyzhang@phy.ecnu.edu.cn}
\affiliation{Quantum Institute for Light and Atoms, School of Physics and Material Science, East China Normal University, Shanghai 200241, P.R. China}
\author{Weiping Zhang}
\affiliation{Department of Physics and Astronomy, Shanghai Jiao Tong University, Shanghai, P.R. China}
\affiliation{Collaborative Innovation Center of Extreme Optics, Shanxi University, Taiyuan, Shanxi 030006, P. R. China}
\author{Pierre Meystre}
\affiliation{Department of Physics and College of Optical Sciences, University of Arizona, Tucson, Arizona 85721, USA}
\affiliation{American Physical Society, 1 Research Road, Ridge, NY 11961, USA}

\begin{abstract}
Hybrid quantum systems can often be described in terms of polaritons. These are quasiparticles formed of superpositions of their constituents, with relative weights  depending on some control parameter in their interaction. In many cases, these constituents are coupled to reservoirs at different temperatures. This suggests a general approach to the realization of polaritonic heat engines where a thermodynamic cycle is realized by tuning this control parameter. Here we discuss what is arguably the simplest such engine, a single qubit coupled to a single photon. We show that this system can  extract work from feeble thermal microwave fields. We also propose a quantum measurement scheme of the work and evaluate its back-action on the operation of the engine.
\end{abstract}


\maketitle

\section{Introduction} 
Experimental advances in single atom and ion manipulation and in nanofabrication have led to an increased interest in quantum thermodynamics, and more specifically in quantum heat engines~(QHE)~\cite{Uzdin2015}. Abah and coworkers proposed~\cite{Abah2012, Ronagel2014} and demonstrated~\cite{Rossnagel2016} a scheme to realize a nanoscale QHE with a single ion. A trapped-ion system was also recently used~\cite{An2014} to carry out an experimental test of the quantum Jarzynski equality~\cite{Jarzynski1997,Mukamel2003}. Other approaches and related fundamental questions in quantum thermodynamics have been considered in systems ranging from quantum degenerate bosonic atoms~\cite{Fialko2012} to superconducting quantum circuits~\cite{Quan2006} and from macroscopically separated quantum-dot conductors coupled to a microwave cavity~\cite{Bergenfeldt2014} to atomic~\cite{Scully2003, Scully2011} or photon gases~\cite{Hardal2015} in optical resonators.

Many hybrid quantum systems can be conveniently described in terms of polaritons. These quasiparticles are quantum superpositions of the system constituents with relative weights that depend on some coupling parameter. The fact that these constituents are typically coupled to reservoirs at different temperatures suggests a general approach to the realization of quantum heat engines where a thermodynamic cycle is realized by periodically varying the control parameter. To an excellent approximation the nature of the quasiparticles is then changed from one to the other of their constituents, so that they are alternatively coupled to one or the other reservoir. 

In previous work~\cite{Zhang2014a,Zhang2014b} we exploited this feature in a phonon polariton based optomechanical QHE \cite{Tercas2016}. Here we expand the same idea to what is arguably the simplest such engine, a single two-state atom or artificial atom (e.g. a superconducting qubit) coupled to a single photon. In this case the polariton modes are the familiar dressed states of quantum optics~\cite{Meystre2007}.  This system could be demonstrated experimentally in a circuit QED environment ~\cite{Walraff2004, Blais2004}.

The paper is organized as follows. Section II establishes our notation and outlines the quantum model of coupled qubit-photon system, including dissipation due to the the coupling of the qubit and the photon to a `cold' and a `warm' reservoir, respectively. Section III describes a quantum Otto cycle based on that qubit-photon system first for the simplest single-photon case, and then the multi-photon and the two-qubit cases. It also derives expressions for the work of the heat engine. Section IV discusses the parameters requirements of the engine working and designs a specific experimental realization based on the circuit QED system. Section V turns to measurement protocols of the work output. It compares explicitly the quantum backaction of dispersive and absorptive measurements on the statistics of the measured results. Finally, Section VI is a summary and outlook.

\begin{figure}[tbp]
\includegraphics[width=7 cm]{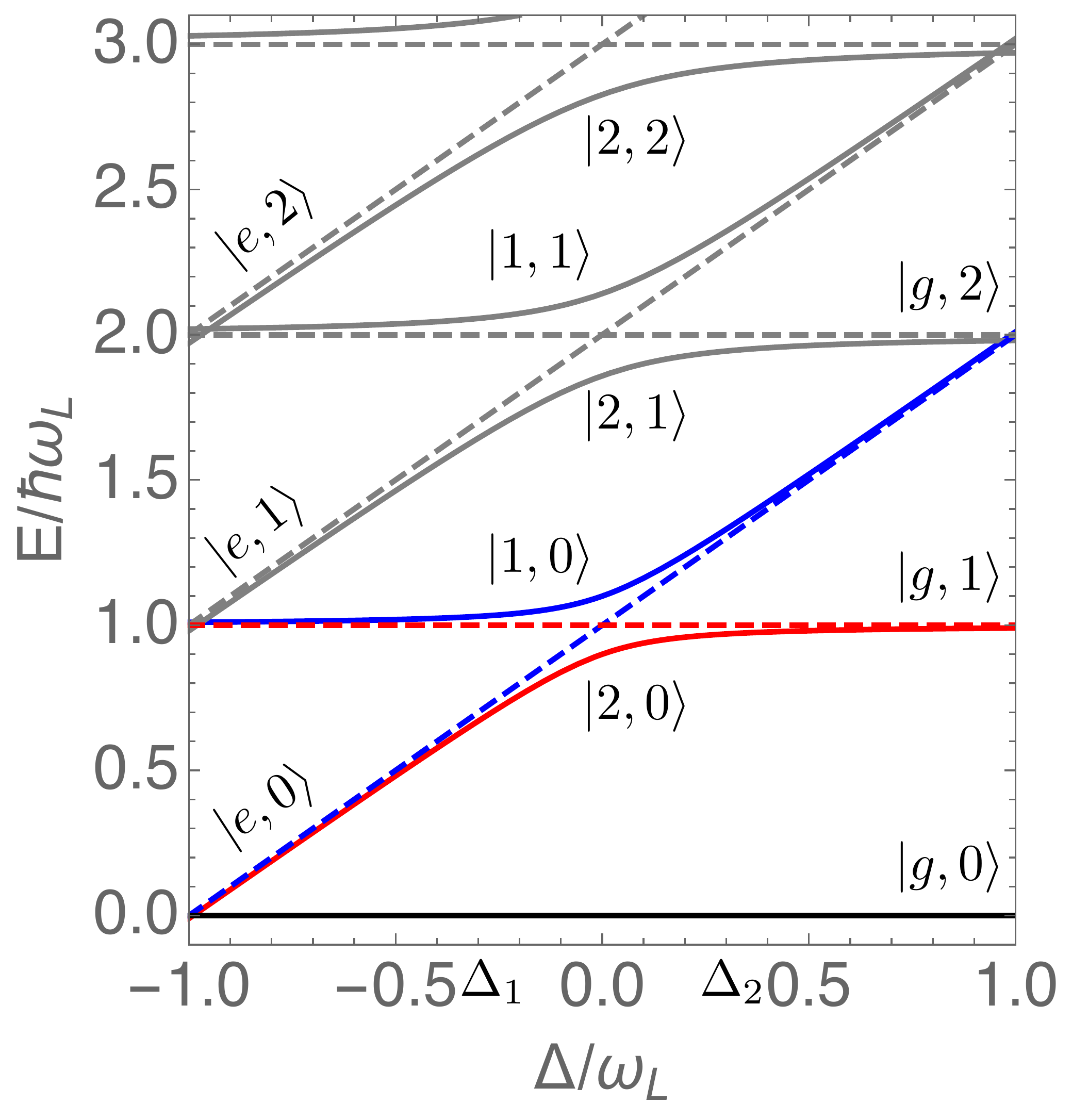}
\caption{Solid lines: dressed qubit energy levels as a function of the qubit-field detuning $\Delta=\omega-\omega_L$ with $\omega$ the qubit transition frequency and $\omega_L$ the cavity field frequency for $g=0.1\omega_L$. Dashed lines: corresponding energy eigenvalues in the absence of interaction, $g=0$. The dressed and bare states are labeled beside the lines. $\Delta_{1,2}$ denote the working frequency range of the quantum heat engine.}
\end{figure}

\section{The cQED System}
We consider a single qubit, which could be either an atom or an artificial atom, trapped inside a high-$Q$ single-mode resonator in a standard cavity QED or circuit QED geometry~\cite{Walraff2004, Blais2004}. In the absence of dissipation and driving and under the rotating wave approximation it is described by the Jaynes-Cummings Hamiltonian
\begin{equation}
H=\frac12 \hbar\omega (\hat \sigma_{z}+1) +\hbar\omega_L
\hat a^{\dagger}\hat a+\hbar g(\hat a \hat \sigma_{+} + {\rm h.c.})\;
\label{H}
\end{equation}
with eigenstates
\begin{eqnarray}
|2,n\rangle = \cos\,\theta_{n}|e,n\rangle-\sin\,\theta_{n}|g,n+1\rangle\;,\nonumber\\
|1,n\rangle = \sin\,\theta_{n}|e,n\rangle+\cos\,\theta_{n}|g,n+1\rangle\;,
\end{eqnarray}
and eigenenergies
\begin{eqnarray}
E_{2,n}&=& \hbar\left[\omega+n\omega_L-\frac{1}{2}({\Omega}_{n}+\Delta)\right]\;,\nonumber\\
E_{1,n}&=&\hbar\left[(n+1)\omega_L+\frac{1}{2}({\Omega}_{n}+\Delta)\right]\;.\qquad
\label{eigenenergies}
\end{eqnarray}
Here $|e\rangle$ and $|g\rangle$ are the excited and ground state of the qubit, with energy separation $\omega$, $|n\rangle$ are Fock states of the field mode of frequency $\omega_L$, $\hat \sigma_i$ are Pauli matrices, $\hat a$ and $\hat a^\dagger$ are bosonic annihilation and creation operators, $g$ is the vacuum Rabi frequency, $\Delta=\omega-\omega_L$ is the qubit-field detuning,  $\Omega_n$ is the quantized generalized Rabi \index{Rabi flopping frequency} frequency
\begin{equation}
{\Omega}_{n}=\sqrt{\Delta^{2}+4g^{2}(n+1)}\;,
\end{equation}
and
\begin{eqnarray}
\cos\,
\theta_{n}=\frac{{\Omega}_{n}-\Delta}{\sqrt{({\Omega}_{n}-\Delta)^{2}+4g^{2}(n+1)}}\;, \nonumber\\
\sin\, \theta_{n}=\frac{2g\sqrt{n+1}}{\sqrt{({\Omega}_{n}-\Delta)^{2}+4g^{2}(n+1)}}\;.
\end{eqnarray}
Fig. 1 shows the first few eigenenergies, illustrating the avoided crossing resulting from the dipole coupling between the qubit and the field at $\Delta = 0$. Importantly for our discussion, the dressed states (qubit-photon polaritons), $|2,n\rangle$ are photon-like for large positive detunings and qubit-like for large negative detunings, and the opposite for the dressed states $|1,n\rangle$. 

The qubit and optical mode are also coupled to thermal reservoirs at temperatures $T_a$ and $T_f$, respectively. In the following we consider the situation where $T_a \approx 0$ and $T_f >0$, a situation that would be characteristic of qubits confined in a cryogenic environment typical of circuit QED experiments and driven by a feeble thermal microwave field. The qubit-field system density operator $\rho$ is therefore governed by the master equation
\begin{equation}
\frac{d\rho}{dt}=-\frac{i}{\hbar} [H,\rho] +\gamma {\cal L}_{\hat \sigma_-} \rho + \kappa (\bar n+1){\cal L}_{\hat a} \rho + \kappa {\bar n}{\cal L}_{\hat a^\dagger} \rho\; ,
\label{master}
\end{equation}
where the Lindblad superoperators are ${\cal L}_{\hat x}[\rho]= \hat x \rho \hat x^\dagger - \frac12 \hat x^\dagger \hat x \rho - \frac 12 \rho \hat x^\dagger \hat x$, $\kappa$ is the cavity mode decay rate, $\gamma$ the qubit spontaneous decay rate, and $\bar n$ the mean number of thermal photons within the resonator bandwidth.

\section{Heat engine}
The difference in  temperatures of thermal reservoirs for the qubit and the photon field allows one to operate a quantum Otto cycle. The following section discusses the operation of a single atom-single photon heat engine that exploits that cycle by varying the detuning $\Delta$.

\subsection{Single-photon case} 
The lowest state $|g,0\rangle$ of the qubit-field system corresponds to the vacuum field state $|0\rangle$ and is therefore qubit-like. The simplest way to operate the qubit-photon heat engine is to limit its operation to the ground state $|g,0\rangle$ and the lowest energy dressed state (lowest polariton branch) $|2,0\rangle$.  By varying the detuning $\Delta$ from a negative to a positive value that state changes its nature from qubit-like to photon-like, thereby changing the thermal coupling from being to a bath at temperature $T_a$ to a bath at $T_f$.  Ideally, in the polariton picture the engine is then {\em effectively} a two-state system, while in the bare-mode picture it actually consist of two coupled qubits like the quantum engine proposed in Ref.~\cite{Linden2010}.

The operation of the engine relies on a four-stroke quantum Otto cycle~\cite{Quan2007}.  The starting point of the cycle is the ground state $|g,0\rangle$ with transition frequency $\omega =\omega_1 < \omega_L$ and corresponding detuning $\Delta_1=\omega_1-\omega_L < 0$, in thermal equilibrium at the qubit reservoir temperature $T_a \approx 0$.  The first, isentropic stroke consists of changing $\omega$ to a new value $\omega_2 > \omega_L$ and detuning $\Delta_2 > 0$. This step can be carried out relatively fast since it does not involve the approach of an avoided crossing where nonadiabatic transitions could be an issue. The second, isochoric stroke is the thermalization of the system with the two thermal reservoirs. Since $T_a \approx 0$ nothing much happens to the qubit constituent of the system, but the field part acquires a finite probability to be excited to Fock states $|n\rangle$ with $n=1, 2,\ldots$ For thermal microwave fields in the 100 GHz range and at  temperatures $T_f$ around 1 K the only state that becomes significantly populated is the Fock state $|n=1\rangle$, with small probability $p_1$. At the end of that step the qubit-field system is then left to a good approximation in the mixed state
\begin{equation}
\rho \approx (1-p_1)|g,0\rangle \langle g,0| + p_1 |2,0\rangle\langle 2,0|. 
\label{rho}
\end{equation}
The ground state component $(1-p_1)|g,0\rangle \langle g,0|$ of $\rho$ plays no active role in the following third, isentropic stroke, so we concentrate of the state $|2,0\rangle$  for now. In that stroke $\omega$ is changed back to its initial value $\omega_1$, and the nature of the dressed state $|2,0\rangle$ changes adiabatically from its approximate photon-like nature, $|2,0\rangle \approx |g,1\rangle$, to its qubit-like form $|2,0\rangle \approx |e,0\rangle$. This step must be carried out slowly enough that nonadiabatic transitions between the dressed states $|2,0\rangle$ and $|1,0\rangle$ remain negligible at the avoided crossing. Finally the last stroke is the spontaneous decay of the qubit-like state $|2,0\rangle$ to the ground state $|g,0\rangle$ at rate $\gamma$.  

The thermalization strokes 2 and 4 are isochoric. Ideally no work is performed on the control field used to change $\omega(t)$ during stroke 1 either, due to the vanishing population on the excited state $|e\rangle$. The only work contribution occurs during stroke 3, a result of the reduction in energy of the excited state population. The average work associated with a full Otto cycle is therefore 
\begin{equation}
W= p_1\left [E_{2,0}(\omega_1)- E_{2,0}(\omega_2) \right ]\,. \label{Wdress}
\end{equation}
It is always negative, i.e. work is produced {\em by} the engine. Noting that $E_{2,0}(\omega_2) \le \hbar\omega_L$ and $E_{2,0}(\omega_1) \le \hbar\omega_1$ we have
\begin{equation}
|W| \le p_1 \hbar |\Delta_1| = p_1 \hbar(\omega_L -\omega_1).
\label{max work}
\end{equation}

We note for completeness that this system can also be operated as a heat pump, provided that $T_a > T_f$, and that the cycle is reversed, with the initial state $|g,0\rangle$ associated with a positive detuning $\Delta$, which is then changed to a negative value in the first stroke. The thermalization of the qubit at $T_a >0$ leads then to a population $p_1$ on the state $|e,0\rangle$. After an adiabatic change of the detuning back to a positive value the photon-like polariton $|2,0\rangle \approx |g,1\rangle$ decays back to the ground state $|g,0\rangle$ at rate $\kappa$. In that mode of operation the average work is equal to $-W$ which is positive, indicating that work is done {\em on} the system. This shows that in case the qubit thermalization dominates the system can operate as a heat pump, but if the thermalization of the field dominates it is a heat engine. 

\subsection{Multi-photon case} 
For $T_f \approx 0$ only the lowest polariton branch $|2,0\rangle$ and the ground state $|g,0\rangle$ are involved in the Otto cycle. In that limit the engine operation is formally identical with that of the optomechanical heat engine~\cite{Zhang2014a}. For larger $T_f$, however, higher polariton branches come into play.  Specifically, at the end of the thermalization stroke $2$ the branches $|2,n\rangle$ with $n \ge 1$ are populated with thermal probabilities  $p_n \approx \bar n^{n+1}/(\bar n+1)^{n+2}$ where $\bar n$ is the average photon number. The result is the appearance of the superposition of several Otto cycles. The average work output $W_n$ produced in stroke 3 by the cycle associated with the polariton mode $|2,n-1\rangle$ is 
\begin{equation}
|W_n| \le \hbar (\omega_L -\omega_1) p_n.
\end{equation}

Following the thermalization stroke 4 the dressed state $|2,n\rangle$ has relaxed to $|1,n-1\rangle$, $n\ge1$. But in contrast to the situation for the $|2,0\rangle$ polariton which is thermalized to the ground state $|g,0\rangle$, the first stroke of the next cycle now costs work. For the symmetric case $|\Delta_1|=\Delta_2$ and perfect adiabaticity that work is precisely equal to the work output of stroke 3 of the previous cycle, and the cycles associated with higher polaritonic modes produce no net work. As shown below the situation is slightly more favorable for the asymmetric situation $|\Delta_1| >| \Delta_2|$, in which case some additional work can be extracted from the engine. For the opposite case $|\Delta_1| <\Delta_2$, in contrast, stroke 1 costs more work than extracted during stroke 3. 


As illustrated in Fig.~1, during the second, isochoric stroke of the engine, the dressed state $|2,n\rangle$ is approximately identical with the bare state $|g,n+1\rangle$ while $|1,n\rangle\approx|e,n\rangle$ provided that $\omega_{2}-\omega_{L}\gg g$. Then at the end of the stroke, owing to the thermalization by an effectively zero-temperature qubit reservoir and a hot microwave reservoir, the state $|1,n\rangle$ is essentially empty, while the ground state $|g,0\rangle$ and the state $|2,n\rangle$ are populated with an approximate microwave thermal distribution, resulting in an average energy
\begin{equation}
\langle H_A\rangle\approx\sum_{n=0}^\infty n\hbar\omega_L p_n,
\end{equation}
where $p_n=\bar n^n/(\bar n +1)^{n+1}$ and $\bar n=1/[\exp(\hbar\omega_{L}/k_{B}T_{f})-1]$ is the mean thermal photon number of the microwave field. During the third, isentropic stroke, $\omega$ is adiabatically changed back to $\omega_1$. Then the dressed states $|2,n\rangle$ and $|1,n\rangle$ approach qubit-excited states $|e,n\rangle$ and photon-excited states $|g,n+1\rangle$, respectively, with their population unchanged, so the average energy at the end is
\begin{equation}
\langle H_B\rangle\approx\sum_{n=1}^\infty \hbar[(n-1)\omega_L+\omega_1 ]p_n.
\end{equation}
The cooling of the heat engine takes place in the fourth, isochoric stroke by coupling it to the qubit reservoir at $T_a\approx 0$. During that stroke the heating from the microwave reservoir remains negligible for $\gamma \gg \kappa$. This results in the transfer of population from the states $|e,n\rangle$ to states $|g,n\rangle$, resulting in the occupation of dressed states $|1,n\rangle$. At the end of that stroke the average energy of the system is therefore
\begin{equation}
\langle H_{C}\rangle\approx\sum_{n=1}^\infty n \hbar\omega_L p_{n+1}.
\end{equation}
Finally, during the following isentropic stroke that brings the qubit frequency back from $\omega_1$ to $\omega_2$ the mean energy becomes
\begin{equation}
\langle H_{D}\rangle\approx\sum_{n=1}^\infty \hbar[(n-1)\omega_L+\omega_2 ]p_{n+1}.
\end{equation}
The work {\em output} is therefore
\begin{equation}
W=\langle H_{B}\rangle-\langle H_{A}\rangle\approx\sum_{n=1}\hbar(\omega_1-\omega_L)p_{n},
\end{equation}
where the sum $\sum_{n=1}^\infty p_{n}$ increases with the raising of temperature $T_f$ with the upper limit $1$ for $T_f \rightarrow \infty$. The maximum work output is therefore equal to the difference between the energy of a single photon and a single qubit, 
\begin{equation}
|W_{\rm max}|=\hbar(\omega_{L}-\omega_{1}).
\end{equation}
The work {\em input}, on the other hand, is 
\begin{equation}
W^{\prime}=\langle H_{D}\rangle-\langle H_{C}\rangle\approx\sum_{n=1}^\infty \hbar(\omega_{2}-\omega_L)p_{n+1},
\end{equation}
So the total work reads
\begin{eqnarray}
W_{\rm tot}&=&W+W^{\prime}\nonumber\\
       &\approx &\hbar(\omega_1-\omega_L)p_1\label{Wtot}\\
       & &+\hbar(\omega_1+\omega_2-2\omega_L)(1-p_0-p_1),\nonumber 
\end{eqnarray}
from which we can find that if $\omega_1$ and $\omega_2$ are chosen to be symmetrically detuned from $\omega_L$,  $\Delta_2\equiv \omega_2-\omega_L=-\Delta_1\equiv \omega_L-\omega_1$, the last term in Eq.~(\ref{Wtot}) vanishes and the total work is precisely equal to the work output in the case of low-temperature microwave reservoir (see Eq. (9)). This is because except for the lowest two dressed states $|g,0\rangle$ and $|2,0\rangle$, the work output and input arisen from the population of the higher dressed states cancel out each other in the Otto cycle, attributing to the symmetric structure of the energy spectrum of the states $|1,n\rangle$ and $|2,n+1\rangle$. Then the total work reaches its maximum value at the precise temperature $T_f$ such that $\bar n=1$ and $p_{1}=0.25$. It then decreases $T_f$ is raised past that point.

There is however nothing fundamental about this result. This upper limit can easily be broken when the condition $\Delta_1 = -\Delta_2$ is no longer imposed. Specifically, for $|\Delta_1|>\Delta_2$ the last term becomes negative so the total work increases (the work output is negative, corresponding to an energy produce by the heat engine), while for the opposite case $|\Delta_1|<\Delta_2$, the total work decreases.  

Finally we note the unique role of $p_1$ in the work performed of the engine. Interestingly, this implies that for some non-equilibrium quantum reservoirs with lowered probability $p_1$, e.g., a squeezed vacuum reservoir with $\rho=\sum_{n=0}^\infty p_{2n} |2n\rangle\langle 2n|$, the total work can be significantly reduced, see Eq.~(\ref{Wtot}).  

\subsection{Two-qubit case}
One can gain some intuition on the origin of the work generated by the engine by observing that provided there is at most one photon in the system the average work $W$ is independent of the number of qubits. Consider for concreteness the case of two qubits.
\begin{figure}[htbp]
\includegraphics[width=7 cm]{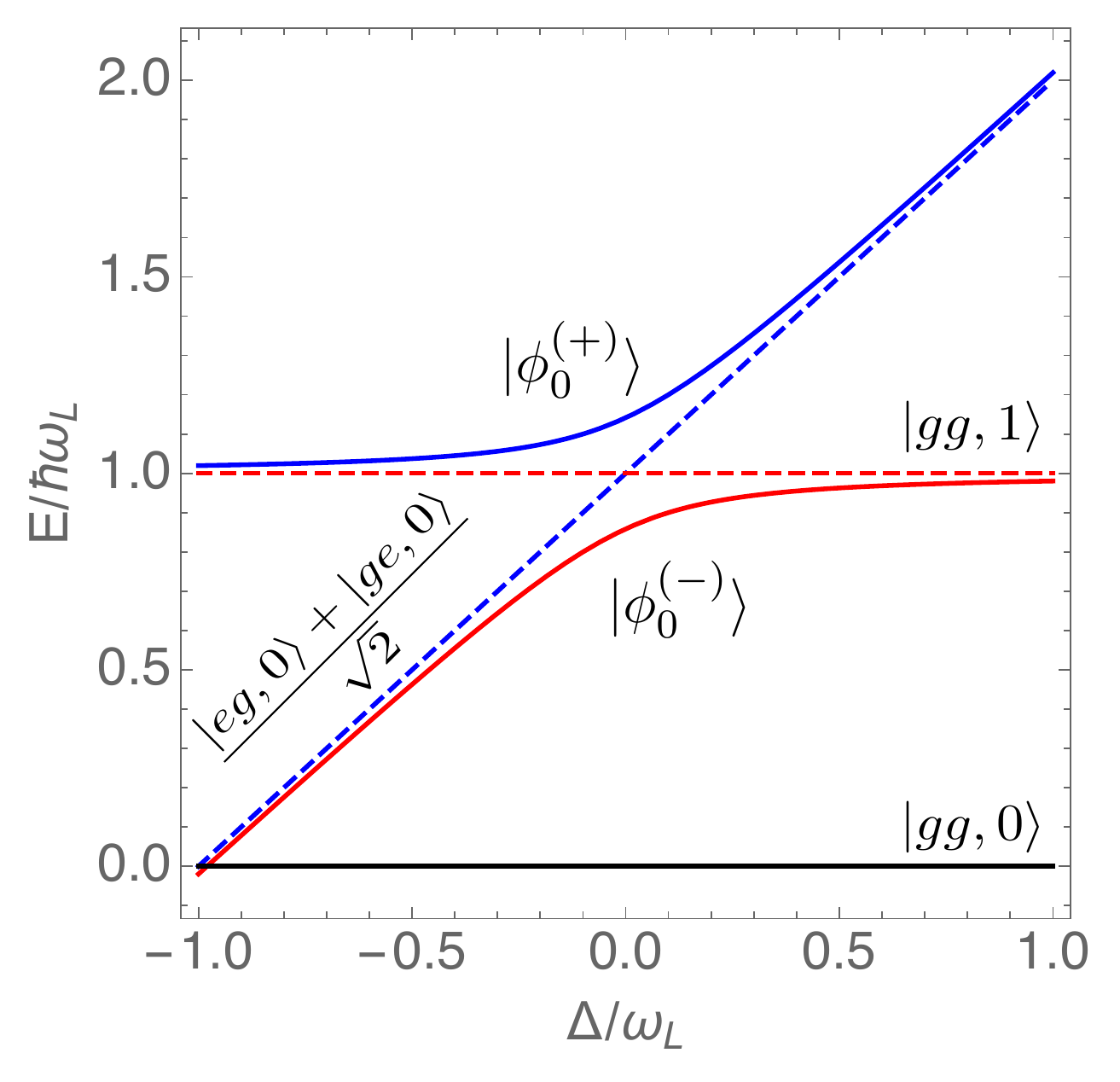}
\caption{Dressed states picture for the two-qubit case. See the text for the definitions of the states $|\phi_0^{(+)} \rangle$ and $|\phi_0^{(-)} \rangle$. Other parameters as in Fig.~1.}
\label{twoatoms}
\end{figure}
The total qubit-field Hamiltonian is then
\begin{equation}
H_2=\frac{1}{2}\hbar\omega (\hat S_z+2)+\hbar\omega_L
\hat a^{\dagger}\hat a+\hbar g(\hat a \hat S_+ + {\rm h.c.})\;,
\end{equation}
where we have introduced the collective spin operators $\hat S_i=\hat \sigma_{1i}+\hat \sigma_{2i}$, $i \in \{z,\pm\}$. Much like for a single qubit, the Hamiltonian can be decomposed into invariant subspaces $H_n$ with $n$ excitations. The subspace $H_0$ is characterized by the ground state $|gg,0\rangle$, and the one excitation subspace spanned by the two dressed states~\cite{Wang2006}
\begin{eqnarray}
|\phi_0^{(+)} \rangle &=&  \sin\frac{\theta}{2}|1,-1\rangle +\cos\frac{\theta}{2} |0,0\rangle, \\
|\phi_0^{(-)} \rangle &=& \cos\frac{\theta}{2}|1,-1\rangle -\sin{\theta}{2} |0,0\rangle\,.
\end{eqnarray}
Here 
\begin{eqnarray}
|1,-1\rangle &=& |gg, 1\rangle,\\
|0,0\rangle &=& \frac{|ge,0\rangle + |eg,0\rangle }{\sqrt{2}},\\
 \cos\frac{\theta}{2} &=& \sqrt{\frac{\Omega_1+\Delta}{2\Omega_1}},\\
  \sin\frac{\theta}{2} &=& \sqrt{\frac{\Omega_1-\Delta}{2\Omega_1}},\\
 \Omega_1 &=& \sqrt{\Delta^2+8g^2} .
 \end{eqnarray}
The corresponding energies are 
\begin{equation}
E_{\phi_0^{(\pm)}}=\frac{\hbar(\omega+\omega_L \pm {\Omega}_1)}{2},
\end{equation}
with energy gap at the avoided crossing between $|\phi_0^{(+)} \rangle$ and $|\phi_0^{(-)} \rangle$ increased to $2\sqrt{2}g$, see Fig.~\ref{twoatoms}. Then for an Otto cycle involving the same sequence of strokes as in the single-qubit case the average work is
\begin{equation}
W=p_1[E_{\phi_0^{(-)}}(\omega_1)-E_{\phi_0^{(-)}}(\omega_2)],
\end{equation}
which is same as Eq.~(8) with $E_{2,0}$ replaced by $E_{\phi_0^{(-)}}$, and is again bounded by Eq.~(9). Having more than one qubit but only one photon doesn't allow one to extract more work from the heat engine, demonstrating that it originates from the photon field. The generalization to $N$ qubits is straightforward, with the energy gap at the avoided crossing increasing to $2\sqrt{N}g$. This larger gap relaxes the time constraints associated with suppression of non-adiabatic transitions.

\section{Experimental considerations}
In the following sections we focus for concreteness on the simplest case of a single qubit coupled to a single photon. Maintaining quantum adiabaticity in the isentropic stroke $3$ requires that changes in the qubit frequency $\omega(t)$ should be slow enough to avoid transitions to the dressed state $|1,0\rangle$, yet faster than the qubit and cavity field decays. Also, $|\omega_2-\omega_1| $ must be much larger than $g$ to guarantee a full photon-like to qubit-like conversion of the nature of the polariton, but the detuning must remain sufficiently small,  $|\Delta|_{1,2} \ll \omega,\, \omega_L$, for the rotating wave approximation and two-level approximation implicit in the Jaynes-Cummings Hamiltonian to remain valid. Turning to the two isochoric thermalization strokes, we note that stroke 2 only necessitates a time long compared to $\kappa^{-1}$, while stroke 4 needs to occur in a time long compared to $\gamma^{-1}$ but short compared to $\kappa^{-1}$ to avoid a significant excitation of 
$|1,0\rangle$. Denoting the duration of the $i^{\rm th}$ stoke as $\tau_i$ the hierarchies of system parameters required for the operation of the heat engine are therefore
\begin{eqnarray}
&&\omega(t), \omega_L \gg |\Delta_{1,2}| \gg g\:, \label{cond1}\\
&&\tau_2 \gg \kappa^{-1} \gg \tau_4 \gg \gamma^{-1}\gg \tau_3 \gg g^{-1}\;.\label{cond2}
\end{eqnarray}

Although these conditions are challenging for traditional cavity QED experiments, they should be realizable in circuit QED devices~\cite{Schmidt2013}. For a resonator frequency $\omega_L\approx 2\pi\times 15$ GHz~\cite{Sandberg2009} the mean photon number $\bar n$ for a thermal blackbody spectrum at  $0.3$K is about $0.1$, the single photon probability $p_1\approx 0.08$, and the occupation probability of the $|n=2\rangle$ state is a negligible $p_2\approx 0.007$. The photon decay rate $\kappa$ of the resonator and the qubit decay rate $\gamma$ can be about $2\pi\times 10$kHz and $2\pi\times 1$MHz, respectively~\cite{Underwood2012}, and the dipole coupling frequency can be as high as $g\approx 2\pi\times 200$MHz~\cite{Peropadre2010,Hoffman2011}, which leaves sufficient time for the adiabatic strokes. 

A specific experimental design of the engine that permits to extract work in an exploitable form, we consider a superconducting transmon qubit. Its frequency can be adjusted by controlling the magnetic flux $\Phi$, with $\omega=\omega_0\sqrt{|\cos(\pi\Phi/\Phi_0)|}$~\cite{Majer2007}. In the bare-mode picture the quantum expression for the infinitesimal average work is
\begin{equation}
	dW=\text{Tr}[\rho \,dH]=\frac{\hbar}{2}(\langle\sigma_z\rangle+1)d\omega.
	\label{dWbar}
\end{equation}
Because $\omega$ is an implicit function of the magnetic field intensity $B$ Eq.~(\ref{dWbar}) is equivalent to $dW=-\mu dB$.  This suggests that the engine can be treated as an artificial magnetic substance with effective average magnetic moment $\mu=-[\hbar(\langle\sigma_z\rangle+1)/2]\partial\omega/\partial B$ located in a circuit loop. In stroke $1$, besides adjusting $\omega$, the change in $B$ also induces a current in the circuit according to Faraday's law, but no work is done by the magnetic substance since $\langle\sigma_z\rangle=-1$ and $\mu=0$. However in the third stroke $\mu\neq0$ so that even when applying an equal change in $B$ the induced current is different. The work outputed by the magnetic substance, i.e. the engine, is responsible for the increase in current, which could be further extracted via coupling to additional elements.

\section{Work measurement} 
A straightforward two-point energy measurement based on Eq.~(\ref{Wdress}) is unsuitable to measure the work of the polariton engine since polaritons are quasiparticles that cannot be directly detected~\cite{Dong2015}. The integral from $\omega_2$ to $\omega_1$ of Eq. (\ref{dWbar}) provides an equivalent expression for the work and suggests that the average work and its fluctuations can be measured by monitoring $\langle\sigma_z\rangle$. However, as should be expected the statistical properties of the measured work depend on the measurement protocol, since different schemes result in general in different measurement backaction. We now compare the results obtained from two different types of measurement. 

\subsection{Dispersive quantum measurements}
If we use a secondary probe beam dispersively coupled to the qubit by the interaction 
\begin{equation}
V_d=\hbar \chi \hat b^\dagger \hat b \,\sigma_z,
\end{equation}
where $\hat b$ and $\hat b^\dagger$ are the annihilation and creation operators of the probe~\cite{Blais2004}, then the repeated homodyne detection of the probe provides a sequence of measurements of $\langle \sigma_z \rangle$.

Ignoring for now effects due to dissipation during the work-producing isentropic stroke $3$, the time evolution of the system is described by the stochastic Schr{\"o}dinger equation~\cite{Jacobs2006,Gambetta2008,Zhang2011}
\begin{eqnarray}  
d|\psi_j\rangle&=&\left [-\frac{i}{\hbar}H-\lambda \left (\hat\sigma_z-\langle \hat\sigma_z\rangle\right )^2\right ]|\psi_j\rangle dt \nonumber \\
&&+\sqrt{2\lambda}(\hat\sigma_z-\langle\hat\sigma_z\rangle)|\psi_j\rangle dw, \label{QT}
\end{eqnarray}
where $\lambda$ characterizes the measurement strength and $dw$ is an infinitesimal Wiener increment. Repeatedly solving Eq.~(\ref{QT}) with the initial state (\ref{rho}) and a slowly-varied $\omega(t)$ generates a set of quantum trajectories $|\psi_j(t)\rangle$.  The mean and variance of the work, as well as the back-action of the measurements are readily obtained from the statistical properties of these trajectories~\cite{Horowitz2012,Hekking2013}.  This also means that the statistics of the work depends on the quantum measurement schemes.

\begin{figure}[hbt]
  \includegraphics[width=7 cm, height=10 cm]{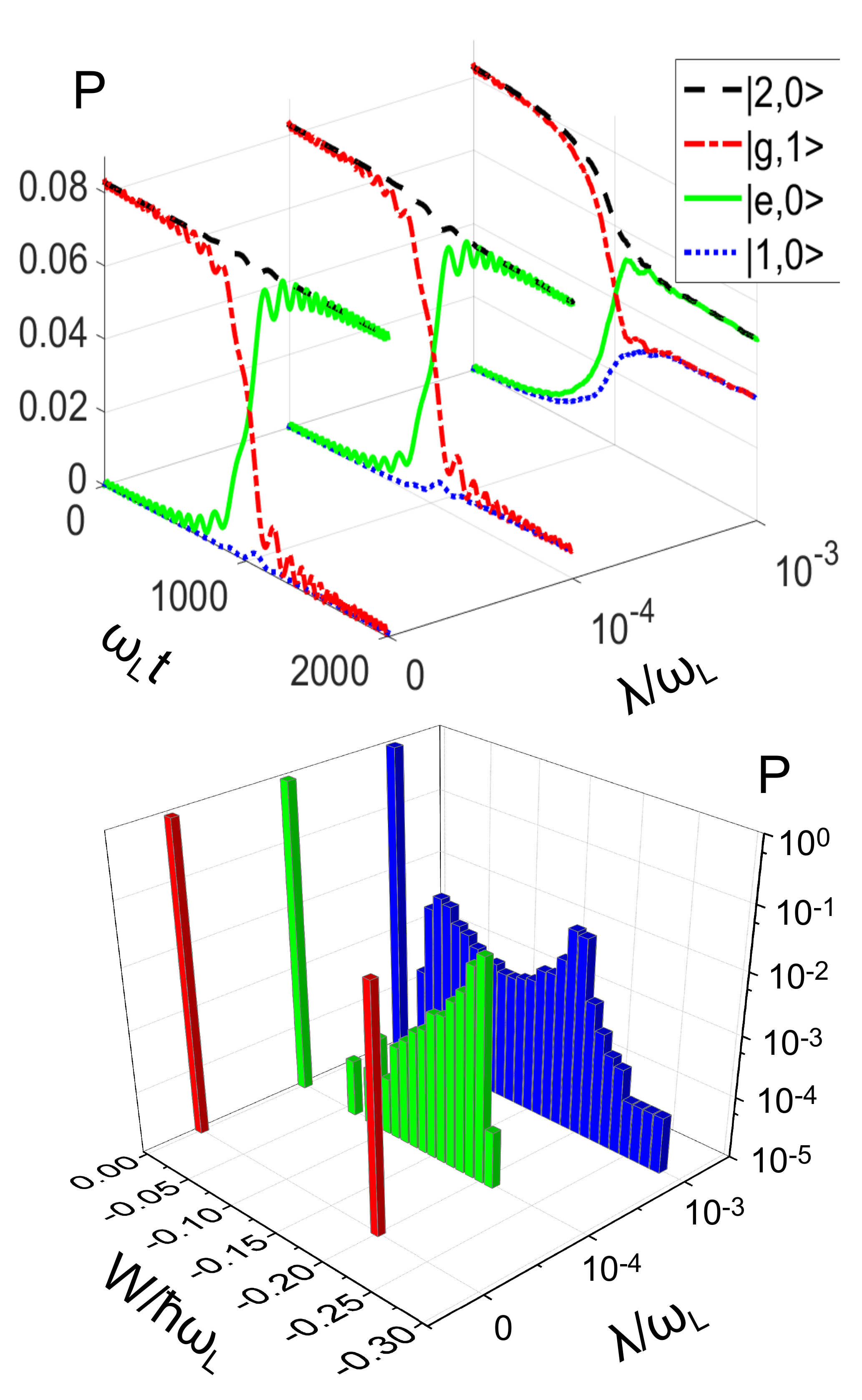}
  \caption{Upper plot: Dressed states and bare states population dynamics during the isentropic stroke 3 in the absence of measurements ($\lambda=0$), and for measurement strengths $\lambda=10^{-4} \omega_L$ and $\lambda=10^{-3} \omega_L$. Here the qubit frequency $\omega(t)$ varies linearly in time from $1.2\,\omega_L$ to $0.8\,\omega_L$ and $g=0.013\,\omega_L$. Lower plot: Log scale probability distribution $P(W)$ of the measured work obtained from 1000 stochastic quantum trajectories for the same measurement strengths.}
  \label{workdistribution}
\end{figure}

Fig.~\ref{workdistribution} summarizes the results of simulations for a set of parameters within state-of-the-art experimental reach. The upper part of the figure plots the evolution of the average populations of the relevant states of the qubit-field system during the third stroke of the cycle, and the lower part shows the probability distribution of the work $P(W)$ extracted in a single cycle of the engine in the absence of measurements and for two measurement strengths $\lambda$. 

In the absence of measurements the bare states $|g,1\rangle$ and $|e,0\rangle$ exchange their populations almost perfectly as $\omega$ is decreased slowly from $\omega_2$ to $\omega_1$ across the avoided crossing. The occupation probability $p_1$ of the dressed state $|2,0\rangle$ remains nearly unchanged, confirming the almost perfect adiabatic conversion of the polariton from photon-like to qubit-like. As expected $P(W)$ is a double-peaked distribution with $W$ taking the value $W=\hbar(\omega_1-\omega_L)$ with probability $p_1$ and $W=0$ with probability $1-p_1$. That latter dominant component~\cite{footnote2} is due to the population $(1-p_1)$ of the state $|g,0\rangle$, which is not involved into the heat engine cycle. 

Because the dispersive coupling $V_d$ of the qubit to the probe field does not commute with the Jaynes-Cummings Hamiltonian (\ref{H}) it couples the two dressed states $|2,0\rangle$ and $|1,0\rangle$ and with an imperfect conversion between the two bare states $|g,1\rangle$ and $|e,0\rangle$. This results in a measurement back-action whereby the peak in $P(W)$ at $W<0$ broadens and spreads towards the zero, as visible in the upper part of Fig.~\ref{workdistribution}. As $\lambda$ increases the adiabatic conversion gradually breaks down and the populations of states $|2,0\rangle$ and $|1,0\rangle$ approach equal values, with the system evolving toward the deterministic steady-state
\begin{equation}
\rho = (1-p_1)|g,0\rangle \langle g,0| + \frac{p_1}{2}(|2,0\rangle\langle 2,0|+|1,0\rangle\langle 1,0|). 
\end{equation}
with a significantly reduced average work.

In addition to measurement-induced dissipation, the effects of qubit and cavity dissipation on the average work during the isentropic stroke $3$ can be evaluated quantitatively by solving Eq.~(\ref{master}). Physically, the spontaneous decay of the qubit from $|e,0\rangle\approx|2,0\rangle$ to $|g,0\rangle$ is dominant for $\omega<\omega_L$, and results in a reduction of the average work. For $\omega>\omega_L$ the thermalization of the cavity mode causes transitions from $|g,0\rangle$ to $|g,1\rangle\approx|2,0\rangle$, increasing the work produced by the QHE. In contrast, for $\omega<\omega_L$ it induces transitions from $|e,0\rangle$ to $|e,1\rangle$, whose population then transfers to the state $|g,1\rangle\approx|1,0\rangle$ during the thermalization stroke $4$, thereby opening up a leak in the Otto cycle. That leak is minimized by imposing $\kappa \ll \gamma$.

\subsection{Absorptive quantum measurements}
\begin{figure}[hbt]
  \includegraphics[width=7 cm, height=10 cm]{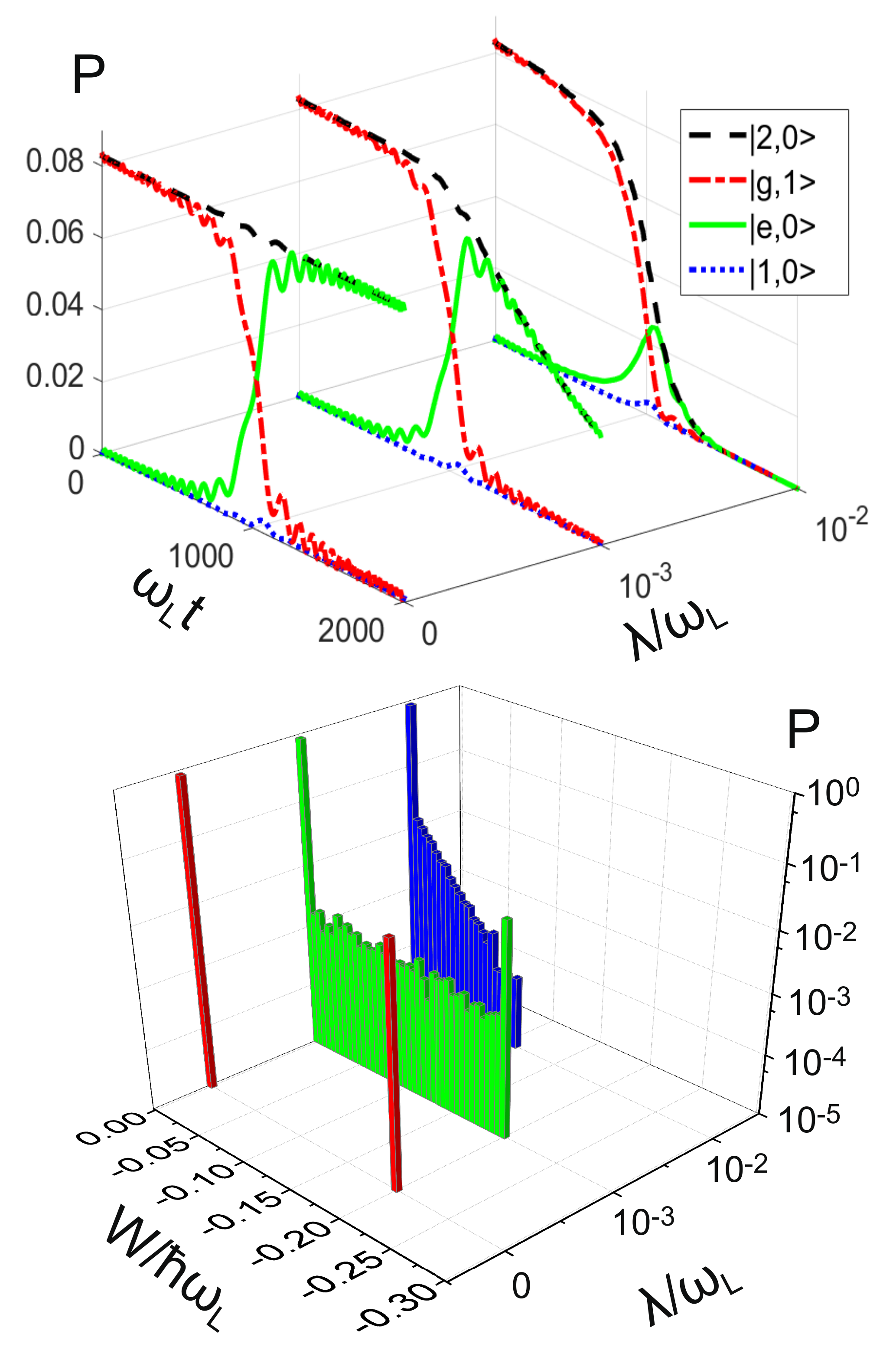}
  \caption{Same plots as Fig.~3 but for the resonant absorptive measurement.}
  \label{workdistribution2}
\end{figure}
To illustrate the dependence of the statistical properties of the measured work on the measurement protocol we compare the homodyne detection with large-detuned probe to a resonant absorption measurement scheme. This can be realized by resonantly coupling a low-density ground state beam of probe two-state systems with the qubit through the interaction
\begin{equation}
	V_a=\hbar\chi(\hat\sigma_+\hat\sigma_-^p+\hat\sigma_-\hat\sigma_+^p),
\end{equation} 
where the superscripts ``$p$'' labels the spin operators of the probe qubits. Ignoring the effects of dissipation during the adiabatic stroke 3 (same as in the dispersive case), the time evolution of the system is then described by the stochastic Schr\"odinger equation~\cite{Jacobs2006},
\begin{eqnarray}
d|\psi_j\rangle &=&[-\frac{i}{\hbar}H+\frac{\lambda}{2} (\langle\hat\sigma_+\hat\sigma_-\rangle -\hat\sigma_+\hat\sigma_-)]|\psi_j\rangle dt \nonumber \\
&+&(\frac{\hat\sigma_-}{\sqrt{\langle\hat\sigma_+\hat\sigma_-\rangle}}-1)|\psi_j\rangle dN,
\end{eqnarray}
where $\lambda$ is a measure of the strength of the measurement and $dN$ is an infinitesimal Ito increment. The results of simulations are shown in Fig.~\ref{workdistribution2} with all parameters as in Fig.~\ref{workdistribution}. In the absence of measurements the evolution of the population of the relevant states and the statistics of the work output are the same in the two figures, as should of course be the case, but as the strength of the measurement increased, the populations of the states $|2,0\rangle$ and $|e,0\rangle$ now decay to zero and the populations of the states $|1,0\rangle$ and $|g,1\rangle$ remain equal to  zero in the end of stroke 3. This is attributed to the absorption from the probe, which results in a measurement-induced energy loss. As a result the average work decreases dramatically, with an associated broadening of the distribution of $P(W)$  towards zero.

\section{Conclusion}
To summarize, we have proposed and analyzed what is arguably the simplest polaritonic QHE, a single qubit coupled to a single photon that operates by absorbing energy from feeble thermal microwave fields. Irrespective of its experimental implementation it offers a straightforward and pedagogically appealing platform for quantitative studies of quantum thermodynamics. Circuit QED realizations of this system seem particularly promising, in which case the work output is readily controllable and extractable, and the influence of the quantum measurement can be demonstrated efficiently. 

\section{Acknowledges}
We acknowledge enlightening discussions with K. Schwab, A. Kamal and L. Zhou. This work is supported by National Key Research and Development Program of China No. 2016YFA0302000. QS and KZ are supported by NSFC Grants No. 11574086, 91436211, 11654005, and the Shanghai Rising-Star Program 16QA1401600. SS and PM are supported by the U.S. Army Research Office. WZ is supported by NSFC Grant No.~11234003.


\begin{thebibliography}{99}
\bibitem{footnote2} Ignoring dressed states with higher coupled system excitations requires that $p_2 \approx 0$, and hence $p_1 \ll 1$.
\end{thebibliography}
\end{document}